\documentclass[a4paper]{jpconf}
\usepackage{graphicx}
\usepackage{xspace}

\def\pT{\ensuremath{p_{\rm T}}\xspace}
\def\kT{\ensuremath{k_{\rm T}}\xspace}
\def\mT{\ensuremath{m_{\rm T}}\xspace}

\begin{document}
\title{Kaon femtoscopy in $\sqrt{s_\mathrm{NN}}$=200 GeV central Au+Au collsions at STAR}

\author{R\'obert V\'ertesi (for the STAR Collaboration)}

\address{Nuclear Physics Institute ASCR, 25068 \v{R}e\v{z}, Czech Republic}

\ead{robert.vertesi@ujf.cas.cz}

\begin{abstract}
Three-dimensional analyses of the pion source revealed a heavy, non-Gaussian tail in the direction of the pair transverse momentum. The interpretation of these pion sources in terms of pure hydrodynamical evolution is, however, complicated by the strong contribution of feed-down from long-lived resonances to the source. On the other hand, kaons provide a much cleaner probe of the expanding fireball. Here we present a recent three-dimensional kaon correlation analysis using Cartesian harmonics decomposition technique. In contrary to pions, the three-dimensional source function of kaons is largely Gaussian. Comparison with thermal simulations and a hydrokinetic model show that resonance decays, as well as non-zero emission duration and/or rescattering in the hadronic phase play an important role. The analysis of the three dimensional  extent of the kaon source w.r.t.\ the pair transverse momentum favors the hydrokinetic model over the exact \mT{}-scaling featured by perfect hydrodynamical models.

\end{abstract}

\section{Introduction}
Bose-Einstein correlations (BEC) between bosons emitted incoherently carry information about the space-time extent of the emitting source. 
Femtoscopy studies the size and evolution of the hot, strongly interacting medium created in high energy heavy ion collisions using BEC measurements. Analysis of two-pion correlation functions with source imaging~\cite{Brown:1997ku} revealed a long-range, non-Gaussian component in the pion source function~\cite{Adler:2006as}. The extraction of the three-dimensional (3D) pion source function using the Cartesian surface-spherical harmonic decomposition technique~\cite{dan05,dan06} in conjunction with model comparisons has permitted the decoupling of the spatio-temporal observable into its spatial and temporal aspects, and the latter into source lifetime and emission duration~\cite{chu08,Chung:2010bb}. However, an interpretation of pion correlations in terms of pure hydrodynamic evolution is complicated by significant contributions from later stages of the reaction, such as decays of long-lived resonances and anomalous diffusion from rescattering~\cite{Vertesi:2007ki,Csanad:2007fr}. A purer probe of the fireball decay can be obtained with kaons, which suffer less contribution from long-lived resonances. The lower yields, however, make kaon measurements more difficult. A one-dimensional kaon source image measurement by PHENIX recently reported an even heavier tail than in pions~\cite{aki09}, but this measurement corresponds to a fairly broad range of the pair transverse momentum $2\kT$, which makes the interpretation complicated. A different aspect of the fireball expansion can be addressed by studying the \kT{}-dependence of the source size. NA49 data at SPS energies, as well as PHENIX data at relatively higher \kT{} values, showed a scaling behavior between pions and kaons, as expected from perfect hydrodynamics models~\cite{Afanasiev:2002fv,aki09}.

\section{Data analysis}

We present recent STAR 3D analyses~\cite{Adamczyk:2013wqm} of the shape and \kT{}-dependent size of the kaon source at mid-rapidity, extracted from BEC using low transverse momentum like--sign kaon pairs in $\sqrt{s_{NN}}$=200 GeV central Au+Au collisions. The kaon source shape was analyzed using 4.6 million 0--20\% central events from 2004, and 16 million 0--20\% central events from 2007. The \kT{}--dependent analysis was carried out using 6.6 million 0--30\% central events from 2004. 
The position of the primary vertex along the beam line was constrained to be $|z|${}$<$30 cm. Kaon tracks with a low transverse momentum (0.1$<$\pT{}$<$1.0 GeV/$c$) were selected in the STAR Time Projection Chamber (TPC)~\cite{stardet}, within a 0.5 T quasi-uniform magnetic field provided by the solenoid surrounding it. The identification of kaons is based on fractional energy loss $\mathrm{d}E/dx$. In the source shape analysis, candidates were required to fulfill the kaon hypothesis to the extent of $|n\sigma_K|<2$ and to be rejected as pions or electrons as $|n\sigma_\pi|>3$ or $|n\sigma_e|>2$, where $n\sigma_X$ is the deviation from the normalized distribution of particle type $X$ at a given momentum. Only pairs with 0.2$<$\kT{}$<$0.36 GeV/$c$ were accepted. In the \kT{}--dependent analysis, kaon pairs were collected in two bins: 0.2$<$\kT{}$<$0.36 GeV/$c$ and 0.36$<$\kT{}$<$0.48 GeV/$c$, with a slightly different kaon selection of $-1.5${}$<${}$n\sigma_K${}$<$2.0 and $-$0.5$<${}$n\sigma_K${}$<$2.0, respectively. Since proton and kaon bands are well separated in both \kT bins, protons are also effectively removed by these cuts.

From an experimental point of view, the 3D correlation function is defined as the ratio of the 3D relative momentum distribution of charged kaon pairs to pairs from mixed events,  
\begin{equation}
C(\mathbf{q}) = \frac{ N_{\rm {same}}(\mathbf{q}) } { N_{\rm {mixed}}
(\mathbf{q}) }, 
\end{equation}
where $2\mathbf{q}$ is the difference between the momenta of the two particles in the pair center-of-mass system (PCMS).
The source function $S(\mathbf{r})$ is the probability of emitting a pair with a spatial separation $\mathbf{r}$. It is linked to the correlation function via the Koonin-Pratt equation~\cite{Koonin:1977fh,lis05,Lednicky:2005tb},
\begin{equation}
C(\mathbf{q})-1 \equiv R(\mathbf{q}) = \int K(\mathbf{q},\mathbf{r}) S(\mathbf{r}) d\mathbf{r},
\label{eq:kp}
\end{equation}
where the kernel can be expressed in terms of the relative wave function,  $K(\mathbf{q},\mathbf{r})=|\phi(\mathbf{q},\mathbf{r})|^{2} -1$, and encodes the quantum correlation as well as final state interactions. 
Source imaging methods numerically invert the above equation without any assumption for the source shape~\cite{Brown:1997ku, dan05}. In the 3D case when the source and correlation functions are handled direction-dependently, the longitudinally co-moving system (LCMS) is often used: The coordinate axes $x$-$y$-$z$ form a right-handed \textit{out-side-long} Cartesian coordinate system oriented so that the $z$-axis is parallel to the beam
direction and $x$ points in the direction of the pair total transverse momentum.
The correlation and source functions can be expanded in Cartesian harmonics basis elements $A^l_{\alpha_1 \ldots \alpha_l} (\Omega_\mathbf{q})$~\cite{dan05,dan06},
where $l=0,1,\ldots$, $\alpha_i=x, y \mbox{ or } z$,  
$\Omega_\mathbf{q}$ ($\Omega_\mathbf{r}$) is the solid 
angle in $\mathbf{q}$ ($\mathbf{r}$) space,
as
\begin{eqnarray}
R(\mathbf{q}) 
=\sum_{l, \alpha_1 \ldots\alpha_l}R^l_{\alpha_1 \ldots \alpha_l}(q) \,A^l_{\alpha_1 \ldots \alpha_l} (\Omega_\mathbf{q})
& \mathrm{\ and\ } &
S(\mathbf{r}) =\sum_{l, \alpha_1 \ldots
\alpha_l}S^l_{\alpha_1 \ldots \alpha_l}(r) A^l_{\alpha_1 \ldots \alpha_l} (\Omega_\mathbf{r }) \ .
\end{eqnarray}

The limited statistics of charged kaons prevented us from carrying out an imaging study. We found, however, that all the moments except for $l$=0 and $l$=2 are consistent with zero, and that the direction independent correlation function R(q) agrees with $R^0$ within statistical errors. Therefore a 3D Gaussian function,
\begin{eqnarray}
  S^G(r_x,r_y,r_z) = \frac{\lambda}{ {(2\sqrt\pi)}^3 R_x R_y R_z}
  \exp[-(\frac{r_x^2}{4 R_x^2} + \frac{r_y^2}{4 R_y^2} + \frac{r_z^2}{4 R_z^2})],
\end{eqnarray} was applied as a trial function for the source, where $R_x, R_y$ and $R_z$ are the characteristic radii of the source in the \textit{out}, \textit{side} and \textit{long} directions, and $\lambda$ represents the overall correlation strength. While $\lambda$ may be sensitive to feed-down from long-lived resonance decays, remainder sample contamination or track splitting/merging not removed by purity and track quality cuts, the radii are virtually independent of these effects. Technically, the fit is carried out as a simultaneous fit on the even independent moments up to $l$=4, yielding $\chi^2/ndf$=1.7 in the source shape analysis, $\chi^2/ndf$=1.1 and $\chi^2/ndf$=1.3 in the 0.2$<$\kT{}$<$0.36 GeV/$c$ and 0.36$<$\kT{}$<$0.48 GeV/$c$ bins in the \kT{}-dependent analysis, respectively. The shape assumption was tested using a double Gaussian trial function and by pushing the fit parameters to the edges of their errors. Other systematic errors were obtained under varying conditions including magnetic field, data collection periods, charge and various sample purity selections. The systematic errors are largely governed by the limited statistics available. 

\section{Results}

\begin{figure}[h]
\begin{minipage}[t]{0.48\textwidth}
\vspace{-12mm}
\includegraphics[width=\columnwidth]{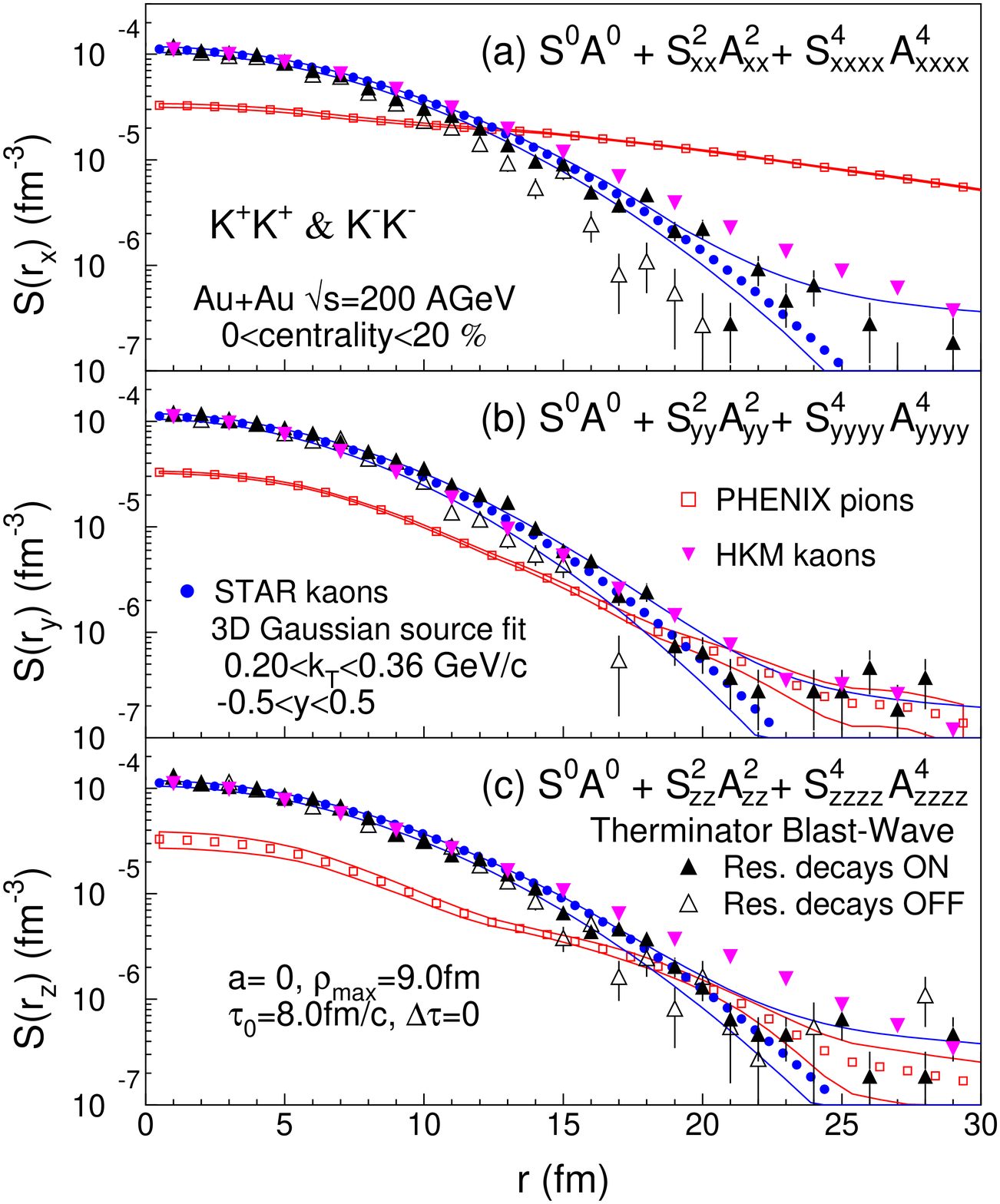}
\vspace{-12mm}
\caption{\label{fig:src}Kaon source function profiles extracted from the data 
    (solid circles) compared to 3D pion source function (squares) from PHENIX
\cite{chu08}, and to the Therminator (triangles pointing upwards) and HKM (triangles pointing downwards) models.}
\end{minipage}\hspace{0.04\textwidth}%
\begin{minipage}[t]{0.48\textwidth}
\vspace{-10mm}
\includegraphics[width=\columnwidth]{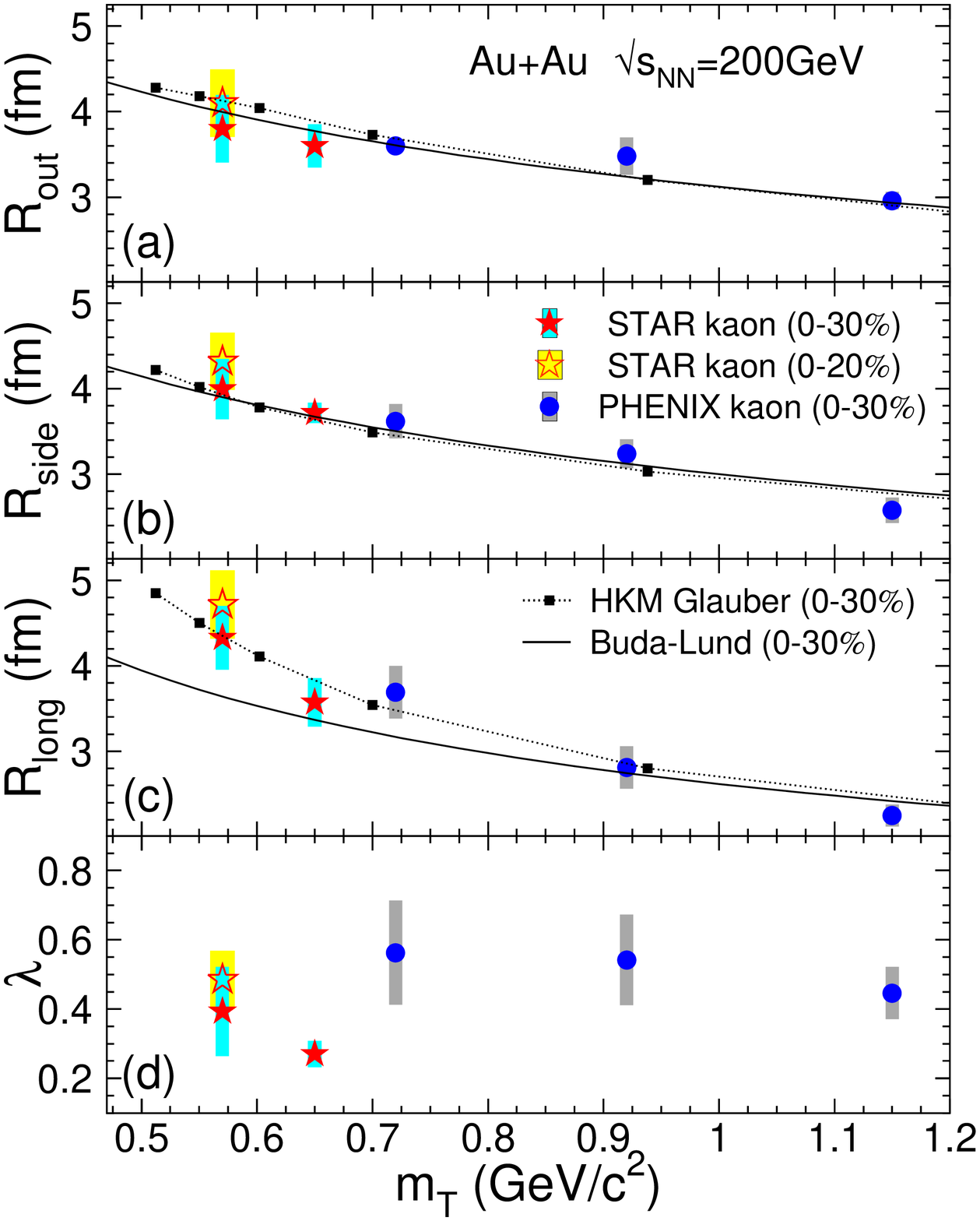}
\vspace{-12mm}
\caption{\label{fig:kt}Transverse mass dependence of Gaussian radii and the $\lambda$ for the 30\% most central Au+Au collisions (solid stars). PHENIX data are also plotted (dots). Squares are HKM, solid curves are Buda-Lund model calculations. The 20\% most central data are also shown for comparison (open stars).}
\end{minipage} 
\end{figure}

The source function profiles in the $x$, $y$ and $z$ directions are shown on 
Figure~\ref{fig:src} (circles). The two solid curves around the Gaussian source function profiles represent the error band arising from the statistical and systematic errors on the 3D Gaussian fit. Note that the latter becomes important for large $r$ values only. The 3D pion source functions from PHENIX~ \cite{chu08} are shown for comparison purposes (squares). While the Gaussian radii are similar, there is a strinking difference between the source shapes of the two particle species, especially in the \textit{out} direction. Note that the PHENIX and STAR pion measurements are fully consistent~\cite{Chung:2010bb}. We have used the STAR tune of the Therminator Blast-Wave model~\cite{kis05,kis06} to gain a better understanding of this difference. The simulation reproduces the source function profiles with emission duration $\Delta\tau$=0 (solid upward-pointing triangles). However, with resonance contribution switched off, Therminator gives a distribution that is narrower than the measurement (empty triangles). Also note that the pion source function is reproduced with Therminator only when non-zero emission duration is assumed~\cite{chu08}.
Recent simulations with the hydrokinetic model (HKM)~\cite{Shapoval:2013bga} show good agreement in the \textit{side} direction, although it is slightly over the measurements at larger radii in \textit{out} and \textit{long} (downward-pointing triangles).

Figure~\ref{fig:kt} shows the dependence of the Gaussian radii in LCMS ($R_{out}$=$R_{x}/\gamma$, ~$R_{side}$=$R_{y}$ and $R_{long}$=$R_{z}$; $\gamma$ is the Lorentz boost in the outward direction from the LCMS to the PCMS frame) as a function of the transverse mass $\mT=(m^{2}+\kT^{2})^{1/2}$. PHENIX kaon data \cite{aki09} are also shown. The perfect fluid hydrodynamics calculations from the Buda-Lund model~\cite{Csanad:2008gt} and the HKM with Glauber initial conditions~\cite{kar10} are plotted for comparison purposes.
While pions are well described by the Buda Lund model in the whole interval shown~\cite{Csanad:2008gt}, low-\kT kaons in the \textit{long} direction seem to favor HKM over the Buda-Lund model, suggesting that the \mT{}-scaling is broken in the \textit{long} direction.

\section{Summary}

We have accounted for the first model-independent extraction of 3D kaon source by the STAR Collaboration~\cite{Adamczyk:2013wqm}, at mid-rapidity in $\sqrt{s_{NN}}$=200~GeV central Au+Au collisions, using low-\kT kaon pair correlations and the Cartesian surface-spherical harmonic decomposition technique. No significant non-Gaussian tail has been observed. Comparison with the Therminator model calculations indicates that, although the transverse extent of the source is similar to pions, the shape and the size in the longitudinal direction is very different. This can be attributed to resonance decays, but also indicates that kaons and pions may be subject to different freeze-out dynamics. Although the Gaussian radii follow \mT{}-scaling in the outward and sideward directions, the scaling appears to be broken in the longitudinal direction. Thus the hydro-kinetic predictions~\cite{kar10} are favored over pure hydrodynamical model calculations. The agreement of HKM simulations with both the source shape and the radii suggests that rescattering in the hadronic phase may also give a non-negligible contribution to the observed source characteristics.


This research was supported in part by the grant CZ.1.07/2.3.00/20.0207 of the ESF programme "Education for Competitiveness Operational Programme", the grant LA09013 of the Ministry of Education of the Czech Republic and by the Hungarian OTKA grant NK 101438.

\end{document}